# Examining Lead-Lag Relationships In-Depth, With Focus On FX Market As Covid-19 Crises Unfolds

**Authors**: Kartikay Gupta (corresponding author), Niladri Chatterjee

**Emails**: maz158144@maths.iitd.ac.in, niladri@maths.iitd.ac.in

Mathematics Department, IIT Delhi, New Delhi, Pin-110016, India

## Highlights

1. A novel technique is proposed to identify the lead-lag relationship between time-series.
2. Comparisons along statistical-significance and forecast-ability, show its superiority over other state-of-the-art models
3. Aligned Correlation measure is proposed, which satisfies most of the metric properties.
4. The technique is used to study the FX market, as the covid-19 epidemic unfolds.

## Abstract

The lead-lag relationship plays a vital role in financial markets. It is the phenomenon where a certain price-series lags behind and partially replicates the movement of leading time-series. The present research proposes a new technique which helps better identify the lead-lag relationship empirically. Apart from better identifying the lead-lag path, the technique also gives a measure for adjudging closeness between financial time-series. Also, the proposed measure is closely related to correlation, and it uses Dynamic Programming technique for finding the optimal lead-lag path. Further, it retains most of the properties of a metric, so much so, it is termed as 'loose metric'. Tests are performed on Synthetic Time Series (STS) with known lead-lag relationship and comparisons are done with other state-of-the-art models on the basis of significance and forecastability. The proposed technique gives the best results in both the tests. It finds paths which are all statistically significant, and its forecasts are closest to the target values. Then, we use the measure to study the topology evolution of the Foreign Exchange market, as the COVID-19 pandemic unfolds. Here, we study the FX currency prices of 29 prominent countries of the world. It is observed that as the crises unfold, all the currencies become strongly interlinked to each other. Also, USA Dollar starts playing even more central role in the FX market. Finally, we mention several other application areas of the proposed technique for designing intelligent systems.

## 1. Introduction

Information plays a critical role in our lives, specifically in financial markets. In financial markets, any accurate information regarding future trends is very financially rewarding. It is sometimes observed that some stock pairs may not have high Pearson correlation coefficient between them, but they are highly correlated at certain lead-lag. Also, it is found that prices of certain financial commodities are following the trends of some other commodity. This phenomenon where a certain time-series replicates the movements of a leading time-series partially at a specific time lag is called lead-lag relationship [1]. The present paper proposes an empirical technique which better identifies the time-varying lead-lag relationship between two time-series.

The rest of the paper is organized as follows. We first give a broad literature review of the lead-lag relationship. Then, we describe the proposed methodology to determine the lead-lag relationship. Then we test this technique to find a known lead-lag relationship in synthetic time-series empirically. Finally, we use this methodology to decipher existing patterns or connections between financial time-series in the FX market.

## 2. Literature Review

It may seem intuitive that any information like a lead-lag relationship which can be utilized for trend discovery should be immediately utilized for making financially profitable transactions. This is not always the case due to factors like delay in information transmission or information arrival. This phenomenon is most phenomenally visible in the time-series of spot and options prices of the same underlying stock or commodity. Here sometimes, it is observed that the options-price series leads the spot-price series [2], which maybe because it is faster to quickly assimilate any new information regarding the future trend into the option price-series as compared to spot price-series. The lead-lag relationship can also be observed in the spot and futures prices of a commodity. The lead-lag relationship in KOSPI200 spot market, its futures market, and its options market are empirically examined and commented upon in the study done by Lee et al. [2]. Tian et al. [3] investigate Taiwan financial markets and find that index-future-prices during non-cash trading-period leads the cash-market during its opening-period. Moews et al. [4] develop an intelligent system to better predict future movements in financial time-series using lagged-correlations with other time-series. Hui et al. [5] find the existence of a time-dependent lead-lag relationship between prices and volume in mini Taiwan exchange futures.

The phenomenon of lead-lag relationship is also observed in the price-series of a commodity being traded at different exchanges. Here, High-Frequency-Trading (HFT) is performed by traders to quickly dissipate any price inconsistencies between two exchanges, while earning huge profits through it. High-Frequency Data in the financial market is gathered at irregular intervals, which makes it challenging to decipher the lead-lag relationship between two different stocks or markets. Thus, an estimator is proposed in [6], which better estimated the cross-covariance by avoiding imputation and using all available transaction. In the study conducted by Robert et al. [7], certain properties of the covariance matrix of increments of two Gaussian processes, partially correlated at some time –lag, is studied.

It may not always be possible to utilize lead-lag information profitably. Still, a lead-lag relationship between two time-series may be indicative of casualty or strong-connection between the two time-series.

Thermal Optimal Path (TOP), first proposed by Zhou and Sornette[8], has been used in the past for obtaining a continuously time-varying lead-lag path between two financial instruments. [8]–[11] are some of the papers which utilise TOP in their analysis of the financial markets. The TOP method has been picked up from physics literature, and it uses Euclidean distance for comparison at the most basic or micro-level. The present measure uses correlation-based distance at micro-level. Correlation distance is more suitable for financial data as it may indicate causality between the two time-series. Further, TOP does not explicitly provide any measure to quantify the strength of the relationship between the two financial time-series.

Another distance measure of significance for any general time-series data is the Dynamic Time Warping (DTW) measure [12], [13]. DTW is generally considered as the best distance measure for time series mining tasks across virtually all domains [14]. DTW measure is especially of advantage in speech recognition [15] where it can decipher the sounds of different words, even when different parts of the word have different elongations. Jin et al. [10] used DTW measure to analyse the network structure of the Foreign Exchange market. Zhu et al. [16] tried to reduce the time complexity of the DTW measure by approximating its value. In work by Silva et al. [14] the effects of relaxing various constraints on the DTW distance measures are studied.

TOP has been one of the most prominent methodologies for empirically finding the lead-lag path. The present methodology shows superior results than TOP. TOP may be considered as a more theoretically evolved version of DTW, as it also uses Dynamic Programming for computation purposes.

The present methodology also improves upon DTW by subtly combining the properties of DTW measure and correlation measure. The proposed Aligned Correlation (AC) can more accurately determine the lead-lag relationship between time-series.

## 3. Proposed Measure

Empirically determination of the best lead-lag path (exact solution) between two time-series requires exponential order of time, as explained in the next section. This is an NP-Hard problem, and Dynamic Programming (DP) is used to obtain an approximate solution in much lesser time. DP is generally used for solving other NP-Hard problems also [17].

The present AC measure takes motivation from the DCCT measure, described in [18]. The AC measure does not require to choose between one of the values of a free parameter 'p', as required in the DCCT measure. Further, the present work provides an in-depth theoretical discussion on the metric properties of the AC measure. It also provides more elaborate testing and comparisons, as compared to the work [18]. Though the DCCT measure is used for profitable pairs trading, the AC measure has been used to study the Foreign Exchange market.

Now, we describe the proposed AC measure in detail. Let $x_t$ , $y_t$ ( $t \in 1,2,\ldots,n$) be two time-series of normalised prices of two stocks. Then the AC's computation requires the use of an alignment path.

Alignment path is a sequence $P = (P_1 , P_2, \ldots P_l \ldots , P_L)$ with $P_l = (p_l , q_l) \in [1{:}n]^2$ for $l \in [1{:}L]$ satisfying the following conditions:

(i) Boundary Condition: Given a parameter psi, (say 100), then $p_1 \leq psi$ , $q_1 \leq psi$, $p_L \geq n - psi$ and $q_L \geq n - psi$ .

(ii) Monotonicity Condition: $p_1 \leq p_2 \leq \cdots \leq p_L$ and $q_1 \leq q_2 \leq \cdots \leq q_L$.

(iii) Step size Condition: $P_{l+1} - P_l \in \{(1,0),(0,1),(1,1)\}$ for $l \in [1{:}L-1]$.

The parameter name 'psi' (which stands for 'Post Suffix Invariant'), has been inspired by [14] where they introduced this parameter to relax the boundary condition in DTW. The alignment path is computed using 'Dynamic programming' techniques as done in DTW measure [14].

The AC measure uses the above definition in its construction. Let us denote the computation of the alignment path by step 1.

## Step 1: Computation of Alignment path

As mentioned earlier, let $x_t$ , $y_t$ ( $t \in 1,2,\ldots,n$) be two time-series of normalised prices of two stocks. Let $rx_t$ , $ry_t$ ( $t \in 1,2,\ldots,n$) denote the consequent return time-series.

Let $CR(p_i, q_i, p)$, a function over the sequence $P_l = (p_l, q_l)$, be defined as follows:

$$rx_{p_j} = x_{p_j+1} - x_{p_j}$$

$$ry_{q_j} = y_{q_j+1} - y_{q_j}$$

$$CR(p_i, q_i, p) = 2 \times \left(1 - \frac{\sum_{j=-p}^{p}(rx_{p_i+j})(ry_{q_i+j})}{\sqrt{\sum_{j=-p}^{p}(rx_{p_i+j})^2}\sqrt{\sum_{j=-p}^{p}(ry_{q_i+j})^2}}\right),$$

Here, 'p' is the window size parameter, which denotes the length of the window. The time-series are appended with $\left\lfloor \frac{p}{2} \right\rfloor$ zeroes at both the ends, so that the above expression can be calculated. In the present research, we use three values of parameter 'p' i.e., 25, 51 and 101.

Then, we find an alignment path $P = (P_1, P_2, \ldots P_l \ldots, P_L)$ which minimises the function $\sum_{i=1}^{L} CR(p_i, q_i, p)$ given the path-constraints as mentioned earlier. This is done through Dynamic Programming techniques as used in DTW measure. This optimization is done in two steps:

1) First, the optimal paths are calculated for each value of the parameter p ( i.e., 25, 51 and 101), which is given by:

$$\text{PATH(p)} = \underset{(p_i,q_i),\, i=1\ldots L}{\text{arg}-\text{min}} \sum_{i=1}^{L} CR(p_i, q_i, p)$$

This is achieved by the DTW algorithm where Euclidean distance is replaced by CR metric, which is also the Euclidean distance between two normalized vectors.

2) Then among these paths, we finally pick the path $P = \{(p_i, q_i),\ i = 1 \ldots L\}$, which minimises the following expression:

$$2 \times \left(1 - \frac{\sum_{i=1}^{L}(rx_{p_i})(ry_{q_i})}{\sqrt{\sum_{i=1}^{L}(rx_{p_i})^2}\sqrt{\sum_{i=1}^{L}(ry_{q_i})^2}}\right)$$

### Step 2: Computation of AC measure

Finally, the AC measure is the correlation-distance along the alignment path, which is calculated as follows:

$$\mathrm{AC} = \underset{(p_i,q_i),\, i=1\ldots L}{Path\ P} \sqrt{2 \times \left(1 - \frac{\sum_{i=1}^{L}(rx_{p_i})(ry_{q_i})}{\sqrt{\sum_{i=1}^{L}(rx_{p_i})^2}\sqrt{\sum_{i=1}^{L}(ry_{q_i})^2}}\right)},$$

where, P is the chosen alignment path ($p_i$, $q_i$).

In the present experiments, the parameter ‘psi’ has been kept equal to the parameter ‘$p$’ just for simplicity.

Here $\sqrt{2(1-\rho)}$ has been chosen as it transforms correlation measure ($\rho$) into Euclidean distance metric between two time-series with unit variance and zero-mean.

# 4. Elaboration of the AC measure

Francisco et al. [19] showed through their work that, in statistical significance, DTW measure satisfies the Triangular Inequality (TI). They reached this conclusion by testing over 15 million triplets for TI, which arose from speech data of 800 time-series. AC measure can also be termed as a ‘loose metric’ as done in [19]. This is because the AC measure is same as the Euclidean distance between normalized time-series along the wrapping path, i.e., DTW measure. Here, the normalizing variance is slightly different from the variance of the original time-series. The slight difference is due to repetition and rare removal of a few terms in the whole time-series. In fact, the AC measure is not equal to the Euclidean metric only because of different alignment of the series along the time. The path obtained here is a valid wrapping path from start to end. It is not the absolute optimal path which minimizes E.D., but it additionally maximises correlation along the path. Many modifications of DTW measure have proposed to put additional constraints on the DTW measure. This measure may also be considered as a measure which puts some additional constraints over DTW measure.

## 4.1 Construction details of the AC measure

It is crucial that CR should be a metric. If we replace it with another measure, we need to make sure that it is a metric. Also, we cannot compare the final $\sum_{i=1}^{L} CR(p_i, q_i, p)$ values for different parameter ‘p’ values. This value $\sum_{i=1}^{L} CR(p_i, q_i, p)$ can be re-arranged to denote the sum of squares of differences between certain normalized segments of the two time-series. This expression is dependent on the parameter ‘p’, and it can not be used for comparison across different ‘p’ values. Thus, instead, we minimize the correlation metric over the aligned path.

### 4.2 Time Complexity

The number of feasible paths grows exponentially with the length of the time-series. The number of feasible paths can be bijectively mapped to the number of paths from top left to bottom right corner in a grid where only right and bottom moves are allowed. The number of such paths is $\binom{2n}{n}$ which is greater than $2^n$. Thus, it requires an exponential order of time to find the exact solution. Here, DP can be used to find an approximate solution quickly.

The proposed AC measure has the same order of time complexity as the DTW measure when seen in terms of the length of the time series. Though, the AC measure has a higher constant term, which increases with the increase in the number of window-sizes (parameter ‘p’) used for finding the alignment path.

## 5. Synthetic Time-Series Experiments

Zhou and Sornette [8] while first introducing Thermal Optimal Path (TOP) [8] method for application in economics, justified its usage with two comparative experiments on synthetic time series data. Here, we will do very similar experiments to determine the superiority of the AC measure path over other models. Another experiment to test self-consistency of TOP results has been conducted earlier several times, like in [9]–[11]. This test will also be performed to determine the validity of the proposed AC technique.

Suppose we have a synthetic time series with a time-varying lead-lag relationship. Then these methods should be able to detect this path, even in the presence of noise experimentally. This experiment is based on the experiments in [8].

In general, the synthetic time-series are of the form:

$$Y(t_1) = aX(t_1 - x(t_1)) + \eta,$$

where $X(t_1)$ is generated through the following process:

$$X(t_2) = bX(t_2 - 1) + \xi,$$

here, the noise-terms $\eta$ and $\xi$ will be explained shortly. Now, given the two time-series $Y(t_1)$ & $X(t_2)$ for some finite length, our aim is to empirically determine the lead-lag structure, i.e., $x(t_1)$.

In the present experiments, we use four different synthetic time series (STS) in two different sets of experiments.

## 5.1 Significance Test

We use STS-1 and STS-2 for testing the significance of lead-lag paths obtained through different algorithms. This test has been earlier employed in [9]–[11], for testing the significance of TOP. The underlying logic of this test is that, if the lead-lag path (x(t)) is significant, then these two synchronized time series, i.e., X(t – x(t)) and Y(t) should exhibit a strong linear dependence. It leads to the following regression.

$$Y(t) = c + aX(t - \langle x(t) \rangle) + \varepsilon(t),$$

In the above equation, the coefficient 'a' should be significantly different from 'zero' for statistically significant dependence. Next we give detail of STS-1.

*Synthetic Time Series 1*

The STS-1 is as follows:

$$Y(i) = \begin{cases} 0.8\,X(i) + \eta, & 1 \leqslant i \leqslant 50, \\ 0.8X(i-5) + \eta, & 50 \leqslant i \leqslant 100, \\ 0.8X(i-10) + \eta, & 101 \leqslant i \leqslant 150, \\ 0.8X(i+10) + \eta, & 151 \leqslant i \leqslant 200, \\ 0.8X(i+5) + \eta, & 201 \leqslant i \leqslant 250, \\ 0.8\,X(i) + \eta, & 251 \leqslant i \leqslant 300, \end{cases}$$

where X(t) is itself a stochastic process given by:

$$X(t_1) = bX(t_1 - 1) + \xi,$$

where b < 1 and the noise $\xi \sim N(0, \sigma_\xi)$ is serially uncorrelated. The factor $f = \sigma_\eta/\sigma_\xi$ quantifies the amount of noise degrading the causal relationship between $X(t_1)$ and $Y(t_2)$. A small f corresponds to a strong causal relationship. A large f implies that $Y(t_2)$ is mostly noise and becomes unrelated to $X(t_1)$ in the limit $f \rightarrow \infty$. Specifically,

$$\text{Var}[X] = \sigma_\xi^2/(1 - b^2)$$

and

$$\text{Var}[Y] = a^2 \,\text{Var}[X] + \sigma_\eta^2 = \sigma_\xi^2 (\frac{a^2}{1-b^2} + f^2) = \sigma_\xi^2 (\frac{a^2 \,\text{Var}[X]}{\sigma_\xi^2} + f^2).$$

All the STS are similar processes with different parameter values.

In our simulations, we generate X and Y of first STS with parameters a = 0.8, b = 0.7, and f = 0.5.

TOP has a free variable ‘Temperature’ which needs to be fixed before finding the path. In work by [11], it is proposed that the temperature value of 2 is generally optimal, and this finding is asserted again in [20]. Here we perform the experiments with the temperature values of 2, 1, 0.5 and 0.2. In the case of AC measure, we use with the following 3 values of parameter ‘p’: 25, 51 and 101, as described earlier.

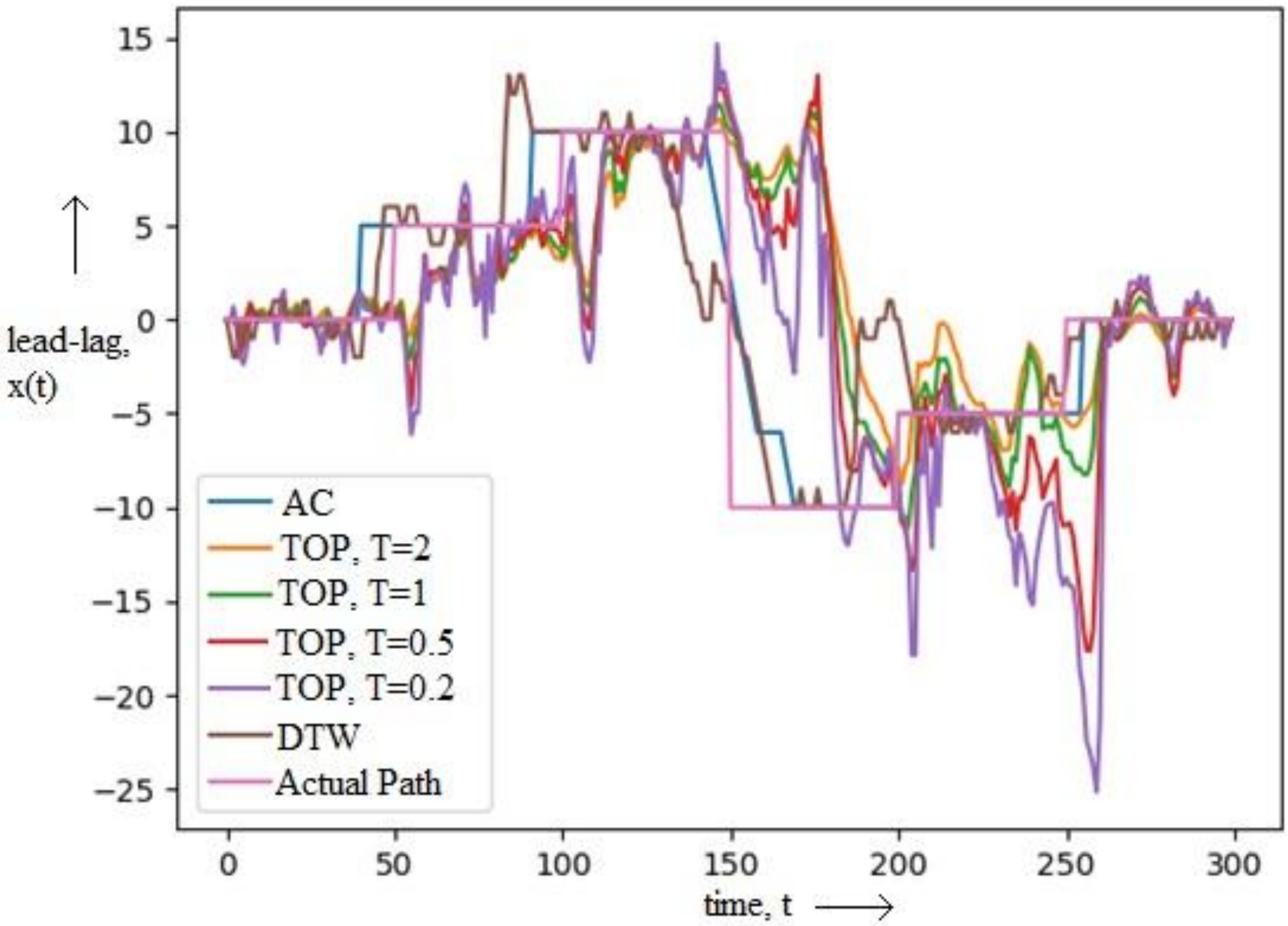

**Figure-1 The Lead-lag Paths:** The empirically determined path x(t) against actual theoretical path by different techniques on STS-1. The path obtained by the AC method is visually more closer to the actual path than other models.

As seen in Figure 1, the AC method is able to perfectly identify the lead-lag structure (x(t)) during the periods when (x(t)) remains temporarily unchanged. It only sometimes fails to capture the path during periods of transitions or jumps. Whereas TOP shows poor performance than AC during both the periods of transitions or no-transitions of x(t). DTW measure has shown visibly better performance than TOP but poor performance than AC. The same phenomenon is observed in Figures 2,3 and 4.

Next is the examination of the empirically obtained lead-lag structure on the basis of the self-consistency test. We perform this test analogously, as described in [11]. We implement this test in moving windows of size 100, which move forward one-time-step from beginning to end of the time series. Thus, we obtain (300 – 100 + 1) such windows over the time series of length 300. Within each window, the two-time-series are synchronized (or not synchronized in the case of 'Unsynched Path'), for estimating the significance of the coefficient 'a'. Table 1 gives the results for this experiment. Here, we observe that among the 201 windows with AC-synchronised time-series, all 201 windows have statistically significant coefficient 'a', at a confidence level of 97.5%. In contrast, for 201 overlapping non-synchronised windows, there are only 61 windows that pass the significant test. When, we observe the mean value of 'a' among windows, where it has a statistically significant non-zero value, we find that the AC's values are close to the actual 'a' values of 0.8, with smaller standard deviation. This indicates the superiority of the AC path over other measures. The model 'Actual Path' denotes the hypothetical model which finds the correct lead-lag path as designed in the STS.

Again, we repeat this experiment on STS-2 and again find that AC has given the best results as seen in Table 2.

| | Model | No. of Windows Significant | Mean a value | Standard Deviation of a values |
|---|---|---|---|---|
| 0 | **AC** | 201 | 0.803827 | 0.094601 |
| 1 | **TOP, T=2** | 103 | 0.473242 | 0.135063 |
| 2 | **TOP, T=1** | 111 | 0.480127 | 0.141182 |
| 3 | **TOP, T=0.5** | 102 | 0.505073 | 0.119492 |
| 4 | **TOP, T=0.2** | 84 | 0.523594 | 0.159355 |
| 5 | **DTW** | 201 | 0.636396 | 0.119881 |
| 6 | **Actual path** | 201 | 0.885287 | 0.055361 |
| 7 | **Unsynched Path** | 61 | 0.474695 | 0.100916 |

**Table-1 Self-Consistency Test Results**: The above table presents the results of the self-consistency test for STS 1. The column 'No. of significant windows' gives the number of windows out of 201, which have a statistically significant non-zero value of the regression coefficient 'a', with a confidence of 97.5%. The columns "Mean value of 'a' " & "Standard Deviation of 'a' " gives the mean and standard deviation of 'a', among windows which have statistically significant non-zero 'a' value.

*Synthetic Time Series 2*

STS-2 is as follows:

$$Y(i) = \begin{cases} 0.8\,X(i) + \eta, & 1 \leqslant i \leqslant 25, \\ 0.8X(i-5) + \eta, & 26 \leqslant i \leqslant 50, \\ 0.8X(i-10) + \eta, & 51 \leqslant i \leqslant 75, \\ 0.8X(i-15) + \eta, & 76 \leqslant i \leqslant 100, \\ 0.8X(i-10) + \eta, & 101 \leqslant i \leqslant 125, \\ 0.8X(i-5) + \eta, & 126 \leqslant i \leqslant 150, \\ 0.8X(i+5) + \eta, & 151 \leqslant i \leqslant 175, \\ 0.8X(i+10) + \eta, & 176 \leqslant i \leqslant 200, \\ 0.8\,X(i+15) + \eta, & 201 \leqslant i \leqslant 225, \\ 0.8\,X(i+10) + \eta, & 226 \leqslant i \leqslant 250, \\ 0.8\,X(i+5) + \eta, & 251 \leqslant i \leqslant 275, \\ 0.8\,X(i) + \eta, & 276 \leqslant i \leqslant 300, \end{cases}$$

where,

$$X(t_1) = bX(t_1 - 1) + \xi,$$

we generate X and Y of size 300 with parameters a = 0.8, b = 0.7, and f = 0.5.

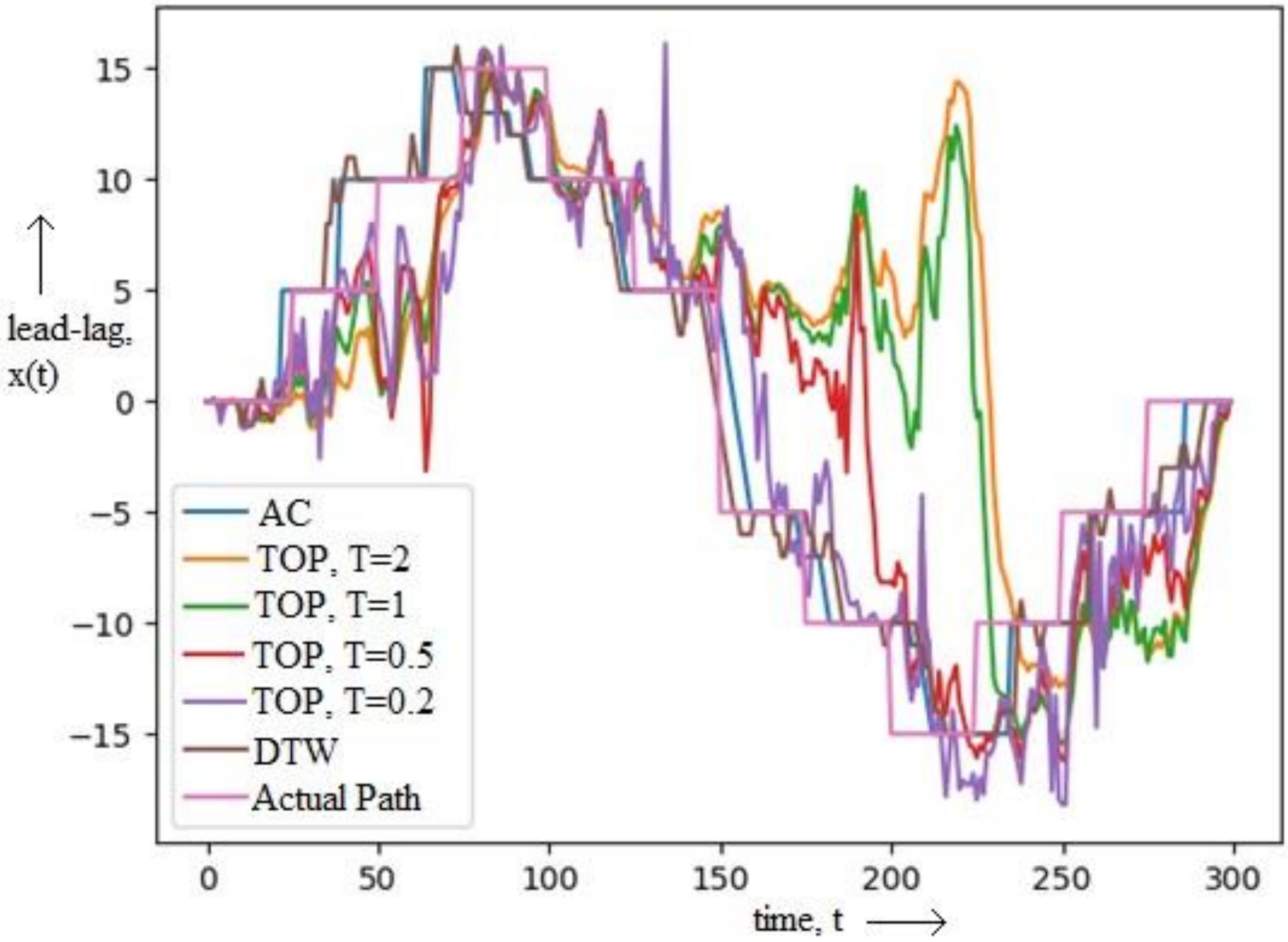

**Figure-2 The Lead-lag Paths:** The empirically determined path x(t) against actual theoretical path by different techniques on STS-2. The path obtained by the AC method is visually more closer to the actual path than other models.

|  | Model | No. of Windows Significant | Mean a value | Standard Deviation of a values |
|---|---|---|---|---|
| 0 | **AC** | 201 | 0.663167 | 0.07755 |
| 1 | **TOP, T=2** | 125 | 0.570674 | 0.166254 |
| 2 | **TOP, T=1** | 124 | 0.466022 | 0.34534 |
| 3 | **TOP, T=0.5** | 131 | 0.565193 | 0.169246 |
| 4 | **TOP, T=0.2** | 147 | 0.519374 | 0.152837 |
| 5 | **DTW** | 201 | 0.570275 | 0.107279 |
| 6 | **Actual path** | 201 | 0.959388 | 0.025072 |
| 7 | **Unsynched Path** | 75 | -0.206854 | 0.157856 |

**Table-2 Self-Consistency Test Results**: The above table presents the results of the self-consistency test for STS 2. The column 'No. of significant windows' gives the number of windows out of 201, which have a statistically significant non-zero value of the regression coefficient 'a', with a confidence of 97.5%. The columns "Mean value of 'a' " & "Standard Deviation of 'a' " gives the mean and standard deviation of 'a', among windows which have statistically significant non-zero 'a' value.

## 5.2 Forecastability Test

Next, we do experiments as done in [8], to test the forecastability of the lead-lag structure. Here, we consider the synthetic time series where X(t) leads Y(t) in general, and thus values of X(t) can be used for predicting future values of Y(t). We use STS-3 and STS-4 for this test. This test examines the ability to obtain correct forecasts through the lead-lag path found by different algorithms empirically. Next, we give details of the STS used in this test.

*Synthetic Time Series 3*

The STS-3 is as follows:

$$Y(i) = \begin{cases} 0.8\,X(i) + \eta, & 1 \leqslant i \leqslant 50, \\ 0.8X(i-5) + \eta, & 50 \leqslant i \leqslant 100, \\ 0.8X(i-10) + \eta, & 101 \leqslant i \leqslant 150, \\ 0.8X(i-15) + \eta, & 151 \leqslant i \leqslant 200, \\ 0.8X(i-10) + \eta, & 201 \leqslant i \leqslant 250, \\ 0.8\,X(i-5) + \eta, & 251 \leqslant i \leqslant 300, \end{cases}$$

where X(t) and $\xi$ are as described earlier.

$$X(t_1) = bX(t_1 - 1) + \xi,$$

we generate X and Y of size 300 with parameters a = 0.8, b = 0.7, and f = 0.5.

We use the present values of X(t) to forecast the future values of Y(t). That is, at each time instance ‘i’, we perform a prediction of Y(i+1), at the time ‘i+1’, which is unknown at the time ‘i’. We first calculate the instantaneous lead-lag time $\tau(i) = \max\{[x(i)], 0\}$ using the lead-lag structure x(i) determined using any of the methods. Here the operator [.], represents the integral part of the number. Now, we obtain the prediction for Y(i+1) as

$$\hat{Y}(i+1) = 0.8X(i+1-\tau(i))$$

In this prediction set-up, we assume that we know the underlying model and the only challenge is to calibrate the lag.

The predicted values $\hat{Y}(i+1)$ are compared to actual values *Y(i+1)*, using Mean Absolute Deviation (MAD) error. MAD is indicative of the maximum loss that will be obtained by financially ‘betting’ on the predicted $\hat{Y}(i+1)$ values.

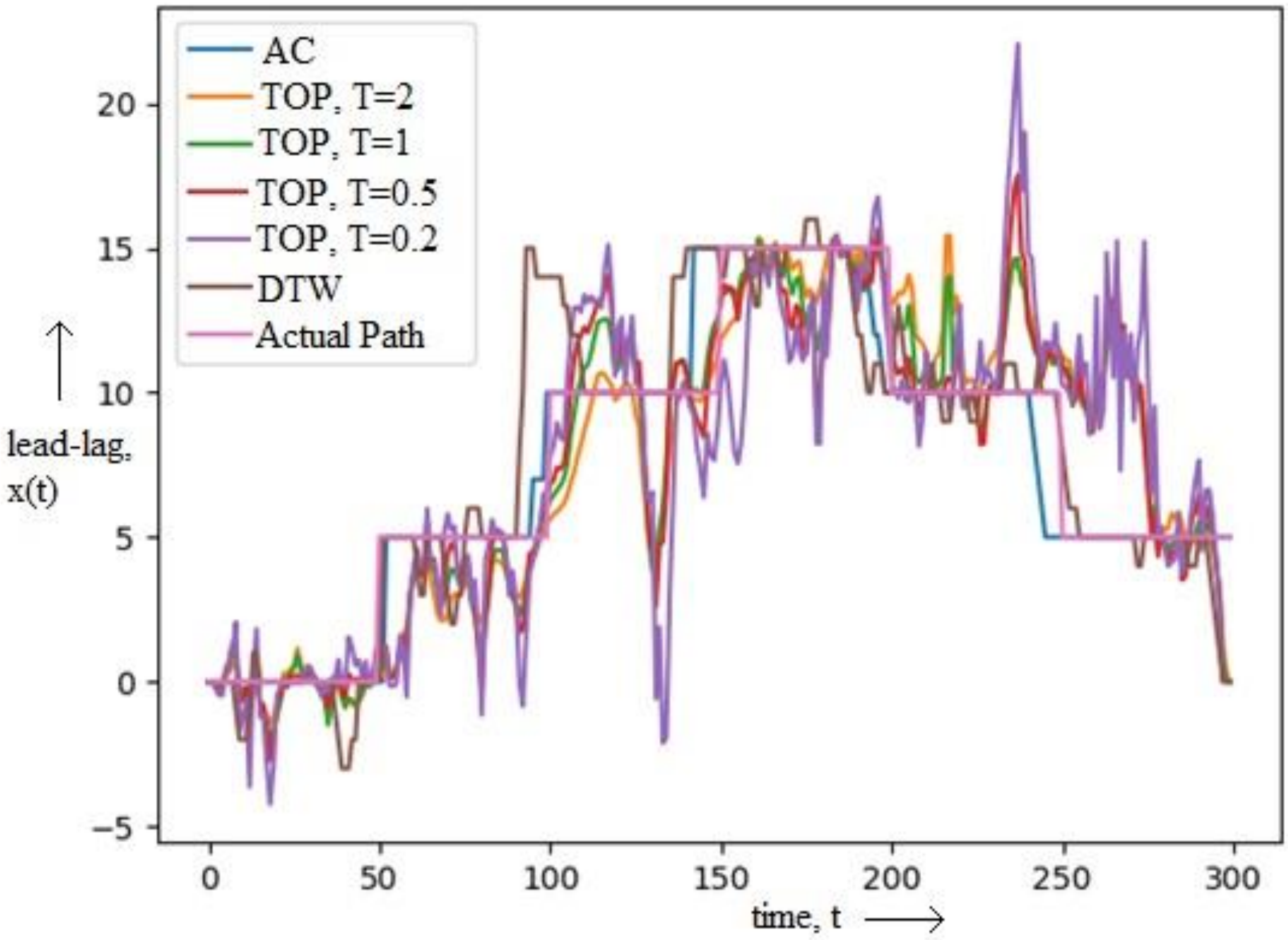

**Figure-3 The Lead-lag Paths:** The empirically determined path x(t) against actual theoretical path by different techniques on STS-3.

| | Model | MAD |
|---|---|---|
| 1 | **AC** | 0.387261 |
| 2 | **TOP, T=2** | 0.581824 |
| 3 | **TOP, T=1** | 0.584978 |
| 4 | **TOP, T=0.5** | 0.58857 |
| 5 | **TOP, T=0.2** | 0.581142 |
| 6 | **DTW** | 0.456977 |
| 7 | **Actual path** | 0.349477 |

**Table 3:** MAD error of the predicted values for different models in STS 3.

*Synthetic Time Series 4*

STS-4 is as follows:

$$Y(i) = \begin{cases} 0.8\,X(i) + \eta, & 1 \leqslant i \leqslant 25, \\ 0.8X(i-5) + \eta, & 26 \leqslant i \leqslant 50, \\ 0.8X(i-10) + \eta, & 51 \leqslant i \leqslant 75, \\ 0.8X(i-15) + \eta, & 76 \leqslant i \leqslant 100, \\ 0.8X(i-20) + \eta, & 101 \leqslant i \leqslant 125, \\ 0.8X(i-25) + \eta, & 126 \leqslant i \leqslant 150, \\ 0.8X(i-30) + \eta, & 151 \leqslant i \leqslant 175, \\ 0.8X(i-25) + \eta, & 176 \leqslant i \leqslant 200, \\ 0.8\,X(i-20) + \eta, & 201 \leqslant i \leqslant 225, \\ 0.8\,X(i-15) + \eta, & 226 \leqslant i \leqslant 250, \\ 0.8\,X(i-10) + \eta, & 251 \leqslant i \leqslant 275, \\ 0.8\,X(i-5) + \eta, & 276 \leqslant i \leqslant 300, \end{cases}$$

$$X(t_1) = bX(t_1 - 1) + \xi,$$

we generate X and Y of size 300 with parameters a = 0.8, b = 0.7, and f = 0.5.

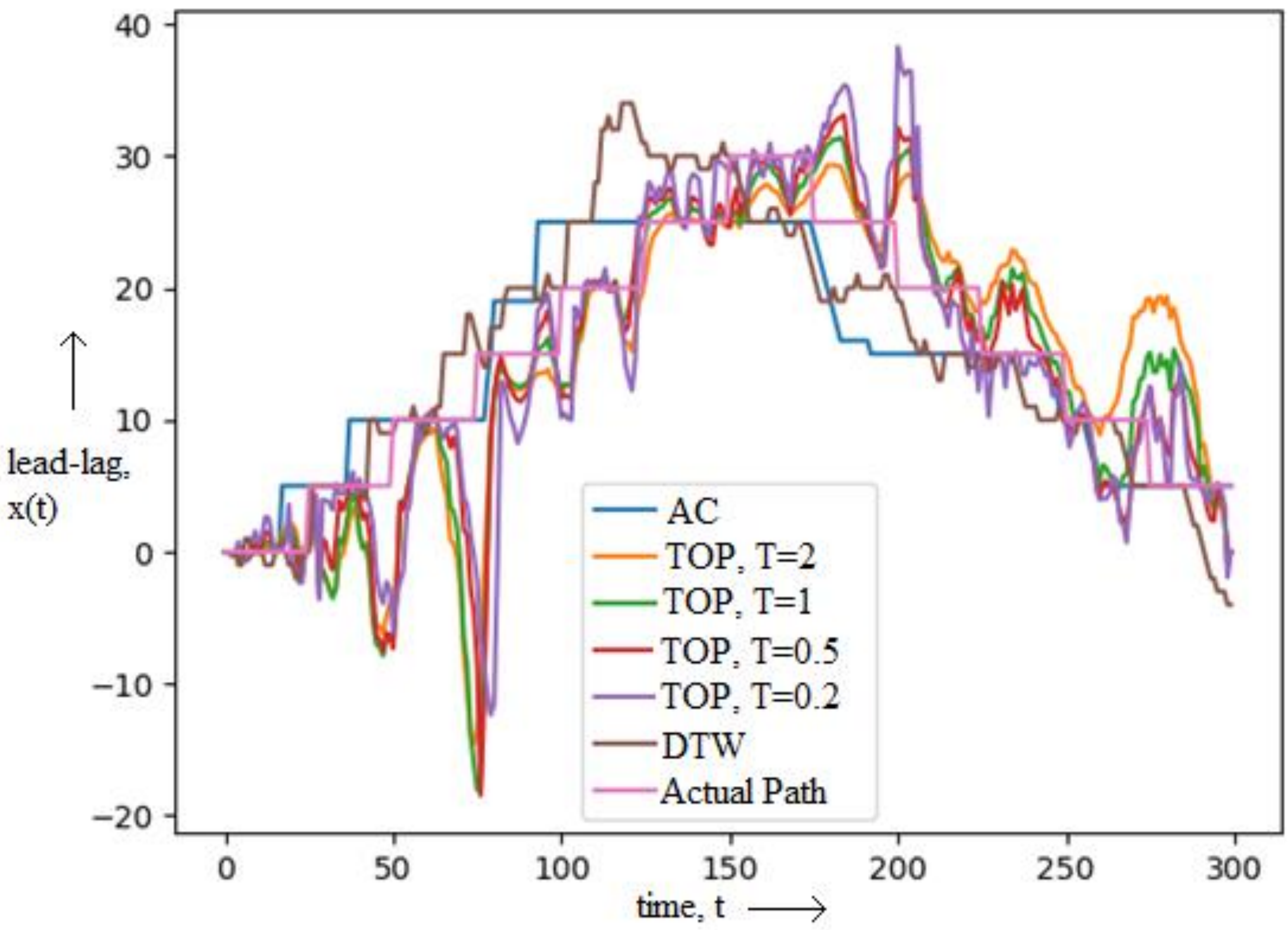

**Figure-4 The Lead-lag Paths:** The empirically determined path x(t) against actual theoretical path by different techniques on STS-4.

| | Model | MAD |
|---|---|---|
| 1 | **AC** | 0.566154 |
| 2 | **TOP, T=2** | 0.835862 |
| 3 | **TOP, T=1** | 0.74202 |
| 4 | **TOP, T=0.5** | 0.674671 |
| 5 | **TOP, T=0.2** | 0.655157 |
| 6 | **DTW** | 0.694811 |
| 7 | **Actual path** | 0.358571 |

**Table 4:** MAD error of the predicted values for different models on STS 4.

Table 3 and 4 give the MAD error of the predictions $\hat{Y}(i+1)$ against actual values $Y(i+1)$ for the different models. We observe that in general, the performance of the AC method is the best except the hypothetical 'Actual Path' model. Thus, though TOP gives better results than a classic-correlation approach, as described in [8], but it performs poorly when compared with the AC approach. Also, DTW measure shows poor performance when compared with AC measure.

## 6. Network-evolution in Foreign Exchange

The present section uses the AC measure to analyse the network evolution of foreign exchange currency as the COVID-19 outbreak unfolds.

Recently, several speculations have been raised regarding the status of the USA dollar (USD) as the world's reserve currency. USD is soon losing its dominance as the world's reserve currency. Many experts are of the view that USD may be replaced by a bucket comprising of RUB, CNY, EUR, oil-backed OPEC currencies etc. Also, due to the recent outbreak of COVID-19 in China, which has first and foremost severely affected the Chinese Stock market, one is tempted to study the topology of correlation networks among major currencies and topology evolution of Foreign Exchange (FX) market.

Topology network analysis through Minimum Spanning Tree (MST) was first introduced in [21], to study the stock prices in financial markets. Jang et al. [22] used topology network analysis to efficiently illustrate the structural and market properties of the financial market. This tool has also been used for financial markets of different regions of the world [23]–[26].

The topology network analysis of the FX market is done in [27], where FX prices of 28 currencies for a period of 12 years from 1990-2002 is analysed. They conclude that USD is the most leading currency in the world. Naylor et al. [28], also used this tool in the FX market and used NZD and USD as numeraries. They found that South-East Asian currencies strongly grouped together during the South-East Asian crisis period. In most of the past such analysis of network evolution, the correlation has been chosen as the preferred metric.

Jin et al. [13] used DTW-measure for doing such analysis of foreign exchange data. DTW measure aligns the two time-series along time, so it can also be used in cases where the two time-series are not of the same length. Also, it is costly and difficult to obtain foreign exchange currency prices of many countries for a long duration, due to different operating hours of exchanges in different countries. DTW measure can be used even if the time-series contain several missing values, without any further data pre-processing step to remove or approximate the missing values. Thus, the DTW measure provides a good alternative to correlation measure as well-argued in [13]. The proposed AC measure also has all these advantages over correlation measure. Further, the AC measure is able to better incorporate the lead-lag relationship effect into its value than DTW measure. It chooses the path along which the correlation is maximum while incorporating any information regarding the lead-lag relationship. As described in [29], [30], the lead-lag relationship may be existing in the foreign exchange market too, so its effect should not be ignored entirely. The proposed AC measure, as we shall see later in the discussion section, mostly chooses the zero-lag path for most of the major currencies, and any deviation is very small. Thus, the proposed AC measure maintains the interpretability of correlation measure while incorporating the effect of any lead-lag relationship.

## 6.1 Data Description

The data consists of foreign exchange currency prices of 29 prominent countries of the world against NZD (see Table 5). The data was sourced from Thomson Reuters Eikon platform. We choose NZD as the numeraire as it was preferred in [13], [28].

| S.No. | Currency Information | Symbol |
|---|---|---|
| 1 | UAE Dinar | AED |
| 2 | Australian Dollar | AUD |
| 3 | Brazilian Real | BRL |
| 4 | BurunFr | BIF |
| 5 | Canadian Dollar | CAD |
| 6 | CongoFr | CDF |
| 7 | SwissFr | CHF |
| 8 | JpYen | JPY |
| 9 | Euro | EUR |
| 10 | GBPound | GBP |
| 11 | New Zealand | USD |
| 12 | IndiaRp | INR |
| 13 | KuwaitDn | KWD |
| 14 | KenyaSh | KES |
| 15 | SriLankaRp | LKR |
| 16 | MynmarKt | MMK |
| 17 | MauritiusRp | MUR |
| 18 | NigeriaNa | NGN |
| 19 | OmanRl | OMR |
| 20 | PakistRp | PKR |
| 21 | ChinaY | CNY |
| 22 | RussiaRub | RUB |
| 23 | Sing$ | SGD |
| 24 | ThaiBaht | THB |
| 25 | TurkLira | TRY |
| 26 | CFAFranc | XAF |
| 27 | SARand | ZAR |
| 28 | KoreaWon | KRW |
| 29 | UgandaSh | UGX |

**Table 5:** 29 prominent currencies used in the analysis of FX market.

The data which is at a frequency of 10-minute, starts at (17-12-2019 16:50) and ends at (16-03-2020 21:00). The time-points which had missing values for any of the currency were dropped from the whole data-set. This data-set can be divided into two parts based on the on-set of the effect of the covid-19 pandemic on the Chinese stock market. The date 03-02-2020 marks the start of the second data-set as on this date, Chinese stocks prices fell drastically, which lead to wiping off around 400 billion USD from the Chinese stock market. Thus, the first part of the data is from (17-12-2019 16:50) to (23-01-2020 21:10), while the second part of the data-set is from (03-02-2020 16:40) to (16-03-2020 21:00).

## 6.2 Methodology

The AC measure has been used for constructing the Minimum Spanning Tree (MST). MST requires the distance measure to be a metric, i.e.; it should satisfy the triangular inequality. As discussed earlier, the proposed AC is a loose-metric. The alignment path in the proposed measure causes the final distance measure to be slightly different from the Euclidean metric (see Section 4 ), which is not generally sufficient enough to violate the triangular inequality. The triangular inequality is more likely to be violated if all the three points lie near to a straight line. This is highly unlikely as the time-series are very high-dimensional points, and further, they have been normalized to lie on the unit circle, during the calculation of the loose-metric. Further, empirically we verify that all the triplets in the two distance matrices, corresponding to two parts of the data, satisfy the triangular inequality. Hence, as done similarly earlier [13], MST based network approach can still be used for the analysis.

## 6.3 Evaluation Criteria

Here, we describe the measures used for evaluating the MST, to study the FX market. We will be using four measures which are as follows.

1) *Mean dissimilarity measure*

   It is defined as

$$L_{\mathrm{MDM}} = \frac{2}{N(N-1)} \sum_{i=1}^{N-1} \sum_{j=i+1}^{N} D_{ij} \,,$$

where D is the NxN dissimilarity matrix, and N is the total number of currencies i.e., 29. The present paper calls the proposed AC measure as a dissimilarity measure as opposed to the term similarity measure (see [13]) as it increases with the increase in farness/dissimilarity between the two objects.

2) *Normalised Tree Length*

It is given by

$$L_{\mathrm{NTL}} = \frac{1}{N-1} \sum_{D_{ij} \in \Theta} D_{ij},$$

where $\Theta$ is the set of edges and it contains the edges present in the MST.

This measure has also been used in [22], to evaluate the MST.

3) *Characterised Path Length*

It is given by

$$L_{\mathrm{CPL}} = \frac{1}{N(N-1)} \sum_{i,j:i \neq j} l_{ij},$$

where $l_{ij}$ is the sum of the weights in the shortest path from node i to node j.

This measure gives the average minimum path distance between any two nodes in the MST [31].

4) *Non-Leaf Nodes*

It is defined as the number of non-leaf nodes present in the graph. This measure helps to judge the loose degree of MST.

### 6.4 Results

**Figure-5 Network 1 corresponding to data-set 1:** The topological network of FX market during the pre-crisis period:

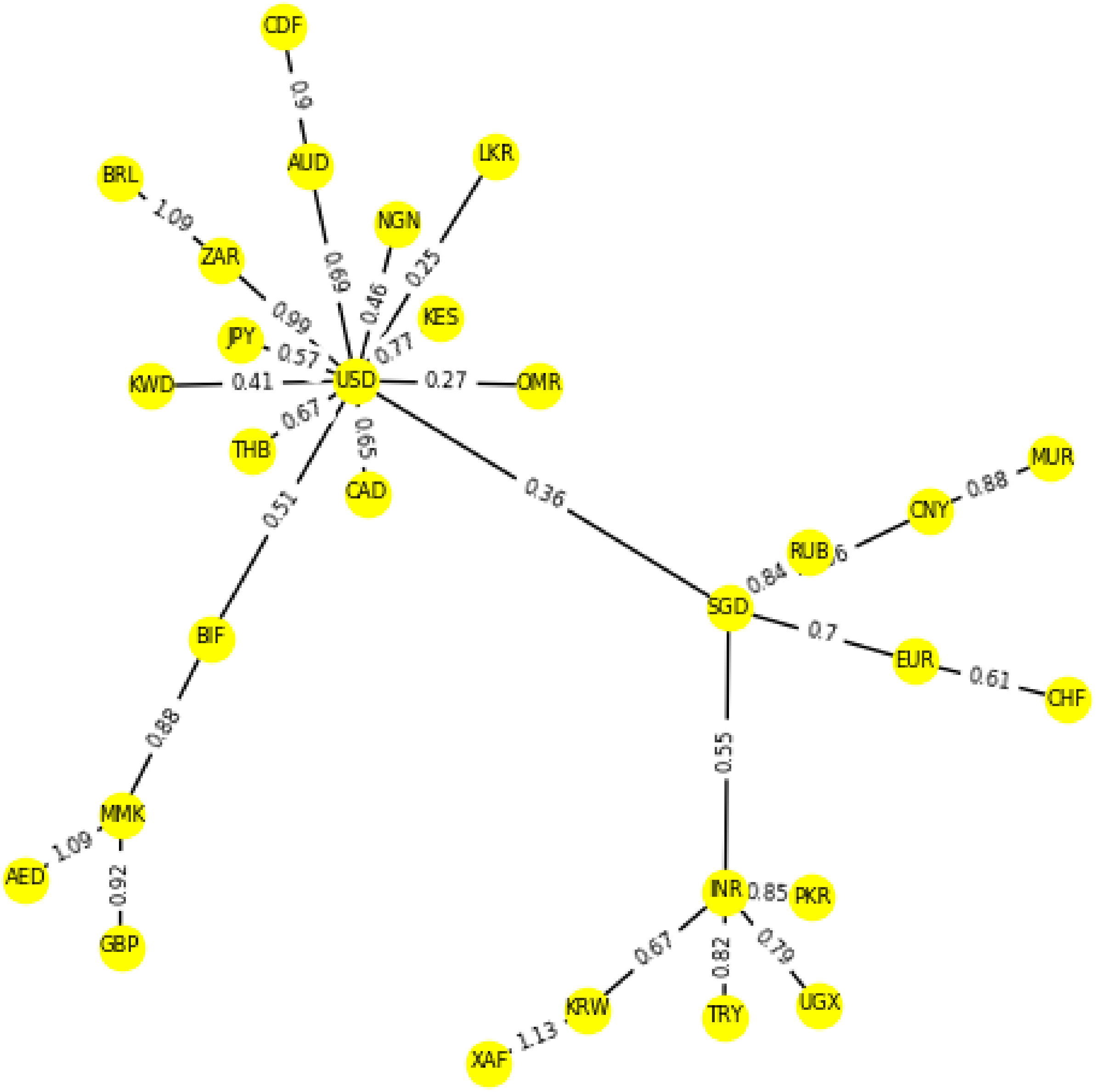

**Figure-6 Network-2 corresponding to data-set 2:** The topological network of FX market during the crisis period:

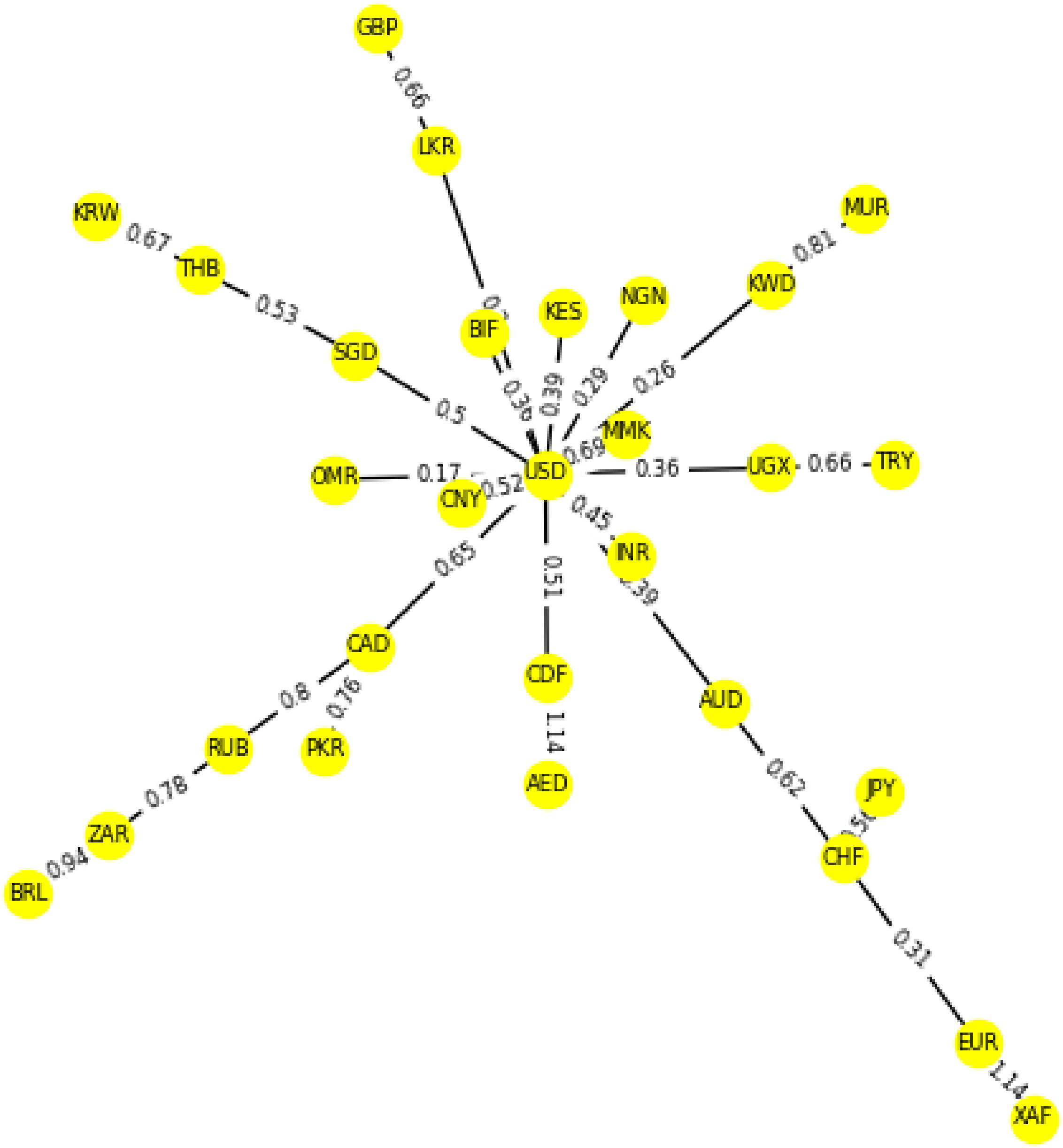

Figures 5 and 6 show the topological network of the FX market during the pre-crisis and the crisis period, respectively. In network 1, three clusters can be seen. The cluster centred around USD is the most prominent one. Another smaller cluster can be seen around the Singapore Dollar (SGD). SGD is closely linked with USD and is also attached to other major currencies like EUR, CNY, RUB, and INR. This point towards the importance of SGD. In earlier studies too [13], SGD has been pointed to be one of the central currencies in the world's FX market. The third cluster is centred around Indian Rupee (INR). The cluster contains regional Asian currencies like INR, PKR and TRY.  The third cluster points towards the emergence of INR as another currency of regional importance. India is one of the fastest-growing major economies in the world. The democratic regime in India has been slowly but steadily liberalizing the economy, which has eased the free flow of goods and services in the region.

As the COVID-19 economic crises began unfolding in China, we see that USD again takes the central role in the FX currencies network of the world. The major currencies like CNY and INR, which were earlier linked to USD through SGD, now attach themselves directly to USD. A tiny cluster can be seen centred around Swiss Franc (CHF) consisting of major currencies like Euro (EUR), Japanese Yen (JPY) and Australian Dollar (AUD). CHF and EUR remain closely linked in both the networks.

| **MST Evaluation Measure** | **Network 1** | **Network 2** |
|---|---|---|
| **Mean Dissimilarity Measure** | 0.962972 | 0.888494 |
| **Normalised Tree Length** | 0.710805 | 0.575337 |
| **Characterised Path Length** | 2.107511 | 1.805699 |
| **Non-leaf Nodes** | 10 | 13 |

**Table 6**: Evaluation measures for the two topological networks.

As observed earlier too [13], the mean dissimilarity measure decreases during the crisis (see Table 6). Essentially the same phenomenon is captured by the two other measures, i.e., NTL and CPL. This denotes that during the crises period, all currencies come closer and start fluctuating synchronously. That is, the movements in the leading currencies are quickly transferred to all the rest of the currencies.

| S.No. | x-axis company | y-axis company | Align-Corr Loose Metric | Aligned Correlation | Correlation Lag=0 | Average Lead/lag | Non-zero ratio |
|---|---|---|---|---|---|---|---|
| 1 | USD | LKR | 0.249 | 0.969 | 0.969 | 0.000 | 0.000 |
| 2 | USD | OMR | 0.271 | 0.963 | 0.963 | 0.000 | 0.000 |
| 3 | SGD | USD | 0.356 | 0.937 | 0.937 | 0.000 | 0.000 |
| 4 | USD | KWD | 0.415 | 0.914 | 0.914 | 0.000 | 0.000 |
| 5 | USD | NGN | 0.462 | 0.893 | 0.893 | 0.000 | 0.000 |
| 6 | USD | BIF | 0.506 | 0.872 | 0.872 | 0.000 | 0.000 |
| 7 | INR | SGD | 0.554 | 0.846 | 0.846 | 0.000 | 0.000 |
| 8 | CNY | SGD | 0.565 | 0.841 | 0.829 | 1.392 | 0.083 |
| 9 | USD | JPY | 0.568 | 0.839 | 0.815 | 0.047 | 0.047 |
| 10 | CHF | EUR | 0.612 | 0.813 | 0.813 | 0.000 | 0.000 |
| 11 | USD | CAD | 0.649 | 0.789 | 0.759 | 0.166 | 0.087 |
| 12 | THB | USD | 0.671 | 0.775 | 0.617 | 0.002 | 0.089 |
| 13 | INR | KRW | 0.672 | 0.774 | 0.753 | 0.156 | 0.136 |
| 14 | AUD | USD | 0.694 | 0.759 | 0.733 | -0.070 | 0.139 |
| 15 | SGD | EUR | 0.701 | 0.755 | 0.744 | -0.028 | 0.032 |
| 16 | USD | KES | 0.767 | 0.706 | 0.706 | 0.000 | 0.000 |
| 17 | INR | UGX | 0.788 | 0.689 | 0.687 | -0.034 | 0.034 |
| 18 | TRY | INR | 0.824 | 0.660 | 0.619 | -0.266 | 0.192 |
| 19 | RUB | SGD | 0.836 | 0.650 | 0.601 | -0.256 | 0.160 |
| 20 | INR | PKR | 0.846 | 0.642 | 0.604 | 0.687 | 0.190 |
| 21 | MMK | BIF | 0.883 | 0.610 | 0.465 | 3.151 | 0.260 |
| 22 | CNY | MUR | 0.883 | 0.610 | 0.428 | -0.955 | 0.401 |
| 23 | CDF | AUD | 0.899 | 0.596 | 0.398 | -0.680 | 0.394 |
| 24 | GBP | MMK | 0.923 | 0.576 | 0.501 | -1.962 | 0.186 |
| 25 | ZAR | USD | 0.994 | 0.507 | 0.492 | -0.100 | 0.100 |
| 26 | BRL | ZAR | 1.089 | 0.407 | 0.312 | -7.301 | 0.388 |
| 27 | MMK | AED | 1.094 | 0.403 | 0.261 | 0.200 | 0.230 |
| 28 | KRW | XAF | 1.132 | 0.359 | 0.096 | 2.621 | 0.744 |

**Table 7(a)**

| S.No. | x-axis company | y-axis company | Align-Corr Loose Metric | Aligned Correlation | Correlation Lag=0 | Average Lead/lag | Non-zero ratio |
|---|---|---|---|---|---|---|---|
| 1 | USD | OMR | 0.167 | 0.986 | 0.986 | 0.000 | 0.000 |
| 2 | USD | LKR | 0.197 | 0.981 | 0.981 | 0.000 | 0.000 |
| 3 | KWD | USD | 0.258 | 0.967 | 0.967 | 0.000 | 0.000 |
| 4 | NGN | USD | 0.294 | 0.957 | 0.957 | 0.000 | 0.000 |
| 5 | CHF | EUR | 0.311 | 0.952 | 0.952 | 0.000 | 0.000 |
| 6 | BIF | USD | 0.356 | 0.937 | 0.937 | 0.000 | 0.000 |
| 7 | UGX | USD | 0.357 | 0.936 | 0.936 | 0.000 | 0.000 |
| 8 | USD | AUD | 0.389 | 0.924 | 0.924 | 0.000 | 0.000 |
| 9 | USD | KES | 0.395 | 0.922 | 0.922 | 0.000 | 0.000 |
| 10 | INR | USD | 0.447 | 0.900 | 0.900 | 0.000 | 0.000 |
| 11 | SGD | USD | 0.496 | 0.877 | 0.877 | 0.000 | 0.000 |
| 12 | USD | CDF | 0.510 | 0.870 | 0.869 | 0.035 | 0.035 |
| 13 | USD | CNY | 0.518 | 0.866 | 0.866 | 0.000 | 0.000 |
| 14 | THB | SGD | 0.527 | 0.861 | 0.841 | 0.447 | 0.115 |
| 15 | JPY | CHF | 0.557 | 0.845 | 0.835 | 0.075 | 0.038 |
| 16 | AUD | CHF | 0.620 | 0.808 | 0.788 | -0.281 | 0.070 |
| 17 | USD | CAD | 0.655 | 0.786 | 0.786 | 0.000 | 0.000 |
| 18 | TRY | UGX | 0.656 | 0.785 | 0.782 | 0.000 | 0.000 |
| 19 | LKR | GBP | 0.662 | 0.781 | 0.741 | 0.065 | 0.085 |
| 20 | THB | KRW | 0.670 | 0.775 | 0.758 | -0.177 | 0.142 |
| 21 | USD | MMK | 0.693 | 0.760 | 0.731 | -0.008 | 0.075 |
| 22 | CAD | PKR | 0.763 | 0.709 | 0.685 | 0.470 | 0.249 |
| 23 | ZAR | RUB | 0.780 | 0.696 | 0.694 | 0.069 | 0.069 |
| 24 | RUB | CAD | 0.803 | 0.678 | 0.451 | -6.133 | 0.308 |
| 25 | KWD | MUR | 0.806 | 0.675 | 0.613 | 0.275 | 0.227 |
| 26 | BRL | ZAR | 0.942 | 0.556 | 0.502 | 1.192 | 0.553 |
| 27 | XAF | EUR | 1.137 | 0.357 | 0.249 | -25.594 | 0.816 |
| 28 | AED | CDF | 1.145 | 0.345 | 0.132 | -18.321 | 0.837 |

**Table 7(b)**

**Table-7(a)(b) Numerical Results:** Table 7(a) abd 7(b) give certain numerical values corresponding to topological networks of data 1 and data 2, respectively. ‘Align-Corr Loose Metric’ gives the value of Aligned correlation loose metric or the aligned correlation measure. Aligned correlation gives the correlation along the aligned path. This value can be similarly interpreted as we interpret the correlation measure. Notice that its value is always greater than the value of correlation measure, as was designed in the proposed technique. ‘Average Lead/lag’ is the average value of lead-lag along the aligned path. ‘Non-zero ratio’ is the ratio of points along the lead-lag path which are different from zero to the total points of the lead-lag path.

As seen in Table 7, the optimal lead-lag path is mostly equal to zero. This is clearly observed in pairs with the lowest distance measure. The lead-lag path sometimes deviates from zero, but the deviations are minor and very rare. In the last few pairs, though the lead-lag path is different from zero, but the correlation values are close to zero. This supports the previous study [30] that it is hard for any significant lead-lag relationship to exist at a frequency of 1 minute or lower in FX markets. But still, the proposed measure takes care of minor corrections, which may have inadvertently arisen into the market prices due to reasons like delay in information-transmission or human-errors.

## 7. Discussion

There exist several studies that can be used for further extending the analysis of correlation-based topological networks. These studies try to overcome the drawback of MST, i.e., loss of information. These studies include [32]–[35], which create graphs which retain more information in them than MST. Since all these techniques employ correlation coefficient, thus there is a possibility of extending the present research in the direction of these techniques.

Other possible areas, where this research may be useful is in the analysis of tick by tick data like in [36]. Tick by tick data needs to be aligned in time, which is done by the proposed measure. Specifically, in the FX market, where there is a vast difference in the liquidity and volumes of different currencies, this time alignment becomes very important. Also, the lead-lag relationship exists substantially in tick-by-tick data. Thus, it will be interesting to see how this research extends on tick-by-tick data.

The proposed technique can also be used in the analysis of the lead-lag relationship of certain other important time-series like done in [10], [11], [20]. The proposed technique, which helps identify the lead-lag relationship, may be extended to help in profitable pairs-trading like successfully accomplished in [37], [38]. It can be used for improving systems which use lead-lag information to achieve better forecasts like [39], [40].

DP is a good tool to solve computationally demanding problems [17]. DP based algorithms are embedded in the chips of computers for computation purposes [41]. Thus, much effort has been put in to reduce the time-complexity of this algorithm further, and now an algorithm is available, which can crudely approximate DTW measure in linear time [42]. The present research can be extended to incorporate these studies.

## 8. Conclusion

DTW and correlation are two of the most frequently used measures in Temporal data-mining literature, and the proposed measure combines them effectively to achieve better task-oriented results. The proposed technique better identifies the lead-lag relationship existing between two time-series. The technique is compared to other models on synthetic time-series data, based on significance and forecast-ability. In terms of significance, the AC measure always finds a statistically-significant path in all the synthetic time-series. In terms of forecast-ability, the forecasts obtained through the proposed technique are closest to the target values. Then we use this technique to explore the topology evolution of 29 prominent FX currencies, as the COVID-19 epidemic unfolds itself. It is observed that after the beginning of the pandemic, USA Dollar assumes a more central position in the world FX market. All the currencies become more closely inter-linked on an average during the crisis period.

## Acknowledgement

The authors were supported by a research grant from IIT Delhi, India.